\documentclass[aps,prb,twocolumn,showpacs,superscriptaddress,10pt]{revtex4-1}
\usepackage{graphicx,amsmath}

\begin{document}

\title{
Two exact properties of the perturbative expansion 
for the two-dimensional electron liquid 
with Rashba or Dresselhaus spin-orbit coupling}

\author{Stefano Chesi}
\affiliation{Department of Physics, Purdue University, West Lafayette, IN 47907, USA}
\affiliation{Department of Physics, University of Basel, 4056 Basel, Switzerland}
\affiliation{Department of Physics, McGill University, Montreal, Quebec, Canada H3A 2T8}

\author{Gabriele F. Giuliani}
\affiliation{Department of Physics, Purdue University,
West Lafayette, IN 47907, USA}

\date{\today}

\begin{abstract}
We have identified two useful exact properties of the perturbative expansion
for the case of a two-dimensional electron liquid with Rashba or Dresselhaus spin-orbit interaction and in the absence of magnetic field.
The results allow us to draw interesting conclusions regarding the dependence of
the exchange and correlation energy and of the quasiparticle properties
on the strength of the spin-orbit coupling which are valid to all orders in 
the electron-electron interaction.
\end{abstract}


\maketitle

\setlength\arraycolsep{0pt}

\section{Introduction}
Modern yearning for understanding the electronic properties of devices made out of
materials in which spins play an important role has rekindled interest 
in the study of the two-dimensional electron liquid in the presence of Rashba or Dresselhaus 
spin-orbit coupling.\cite{bychkov84a,bychkov84b,dresselhaus55}
To date, the effects of the electron-electron
interactions in these system have been studied by means
of approximated methods, notably the random-phase-approximation (RPA) and the mean-field theory.\cite{chen99,saraga05,nechaev09b,nechaev10,Agarwal2010,ChesiPhD,chesi07b,chesi07d,Chesi2010,simion10,Abedinpour2010} 
Quasiparticles parameters were studied in Refs.~\onlinecite{chen99,saraga05,nechaev09b,nechaev10}
while Refs.~\onlinecite{Agarwal2010, chesi07b,chesi07d,Chesi2010,simion10,ChesiPhD,Abedinpour2010} focused on the ground-state properties
and mean-field phase diagram. Recent studies of spatially inhomogeneous
Overhauser's type of instabilities (chirality, charge and spin) and of the spin-susceptibility
in these systems can be found in Refs.~\onlinecite{simion10} and \onlinecite{ChesiPhD,zak10,Agarwal2010}, respectively (including non-analytic corrections\cite{zak10}). Calculations of realistic systems of lower dimensionality in the presence 
of spin-orbit coupling have also been carried out by means of numerical density 
functional theory methods.\cite{valinrodriguez02a,valinrodriguez02b}  
However, for want of more accurate exchange and correlation functionals,
the latter have been implemented by employing formulas for the energy of the electron liquid 
obtained in the absence of spin-orbit coupling, an approximation that a priori 
could appear at best rather crude. Only recently, quantum Monte-Carlo numerical data 
of the energy were obtained for electrons in two dimensions with Rashba spin-orbit coupling.\cite{ambrosetti09}

In this paper we analyze some formal aspects of the perturbative expansion in the 
electron-electron interaction, without magnetic field and under the assumption that the system will behave 
as a Fermi liquid. In doing so we will show that certain exact identities lead to the 
conclusion that, at least in the high density regime or for small spin-orbit coupling strength,
the effects of the interactions are only marginally influenced by the spin-orbit coupling.
In particular we will show that, to all order in the electron-electron interaction,
the corrections to the total energy and the Fermi liquid properties formally vanish to lowest 
order in the spin-orbit coupling. As consequence, our results lend some degree of 
justification and theoretical underpinning for many of the numerical studies carried out 
by neglecting such effects. 

In Ref.~\onlinecite{aleiner01} a canonical transformation was proposed for the case of small dots
that allows one to renormalize the hamiltonian into one in which a generic combination 
of Rashba and Dresselhaus spin-orbit coupling terms only appears in second order. The vanishing of
such spin-orbit interaction effects at small coupling was later verified within the RPA in 
Ref.~\onlinecite{saraga05} for the self-energy. Our analysis not only extends these results to the fully interacting case 
but also explicitly shows how the repopulation of momentum space with respect to 
the non interacting situation\cite{chesi07b} is responsible for the renormalization of the Rashba 
and Dresselhaus coupling. 
Our formalism also sets the stage for a systematic analysis 
of the perturbation theory to all order for regimes in which the series is expected to converge. 
The specific application to the case of Coulomb interaction in the high-density regime
and a comparison of analytic formulas for the exchange and correlation energy 
to the numerical Monte-Carlo results of Ref.~\onlinecite{ambrosetti09} can be found in Ref.~\onlinecite{Chesi2010}. 

The paper is organized as follows: In Section \ref{basic_formalism} we provide the definition of
the model and the necessary notation; In Section \ref{energy_expansion} we prove an 
exact identity concerning the dependence of the total interacting energy on the suitably defined 
density dependent bare spin-orbit coupling and the chirality, the relevant Fermi 
surface repopulation parameter; In Section \ref{self_energy} a similar relationship is obtained for
the self-energy perturbative series; Finally, Section \ref{discussion} provides a 
discussion of our results.
\section{Generalities and basic formalism}\label{basic_formalism}
Consider the many-body problem associated with the hamiltonian
\begin{equation}
\label{Hint}
\hat {H} ~=~ 
\sum_i \left[ \frac{ \hat{\bf p}_i^2}{2m}+\alpha \,(\hat{\sigma}_{xi} \hat{p}_{yi}-
\hat{\sigma}_{yi} \hat{p}_{xi} ) \right]
+ \frac{1}{2}\sum_{i\neq j}
\frac{e^2}{| \hat{\bf r}_i- \hat{\bf r}_j|}  ~,
\end{equation}
where the presence of an homogeneous neutralizing background is understood. The second term in the square brackets represents a spin-orbit coupling of the linear Rashba type and can be readily 
seen to be equivalent to a Dresselhaus spin-orbit coupling term.\cite{bychkov84a,bychkov84b,dresselhaus55}
While in Eq.~(\ref{Hint}) the interaction is Coulombic, as it will be shown, our results are valid for a generic interparticle potential.
The single particle solutions of the non interacting system are the well known chiral 
states: 
\begin{equation}
\label{phi0kpm}
\varphi_{ {\bf k} , \pm} ( {\bf r} )  =
\frac{e^{i {\bf k} \cdot {\bf r} }}{\sqrt{2L^2}}
\left(
\begin{array}{c}
\pm 1\\
i e^{i\phi_{\bf k}}
\end{array}
\right)
\equiv
\frac{e^{i {\bf k} \cdot {\bf r} }}{\sqrt{L^2}}
| { \bf k } \pm \rangle ~
\end{equation}
where $L$ is the linear size of the system, $\phi_{\bf k}$ is the angle between
the wave vector ${\bf k}$ and the $x$ axis and we have defined the spinors $| { \bf k } \pm \rangle$.

It is useful at this point to simplify the notation by rescaling all wave vectors by
$k_F=\sqrt{2 \pi n}$, with $n$ the areal density, and all energies (frequencies) 
by $\epsilon_F = \frac{\hbar^2 k_F^2}{2m}$ the Fermi energy in the absence of spin-orbit coupling 
($\frac{\epsilon_F}{\hbar}$). With the present notation the eigenenergies corresponding 
to~(\ref{phi0kpm}) can be simply expressed as
\begin{equation}
\label{epsilon0k2}
\epsilon_\pm(k) =  k^2  \-\mp g \, k   ~,
\end{equation}
where we have defined the dimensionless coupling constant
\begin{equation}
\label{g_def}
g ~=~ \frac{2 m \alpha}{\hbar k_F} ~,
\end{equation}
which plays an important role in the following. Without loss of generality, we assume $g\geq 0$. 

As discussed in Ref.~\onlinecite{chesi07d} the unperturbed, spatially uniform states with 
symmetric k-space occupation can be completely characterized by the generalized 
chirality $\chi$. This parameter is defined by the Fermi surfaces, two circles 
with radii given by
\begin{equation}\label{kpm}
k_\pm=\sqrt{|1 \pm \chi|} ~.
\end{equation}
When both chiral bands are occupied this quantity coincides with the regular chirality: 
\begin{equation}
\label{chi_def}
\chi=\frac{N_+-N_-}{N_++N_-} < 1,
\end{equation}
where $N_\pm$ is the number of electrons in each chiral band.
This is the case for not too low densities and it is the situation we will deal with here.\cite{comment_genchi}

The corresponding occupation numbers 
can then be expressed as follows:
\begin{eqnarray}\label{n_plus_minus}
n_\pm(k)=\theta(\sqrt{1\pm \chi}- k) ~~~,    &   \quad {\rm for }~0 \leq\chi < 1   ~,
\end{eqnarray}
where $\theta(x)$ is the usual step function.
The non interacting energy per particle (in units of $\epsilon_F$) 
is in turn given by
\begin{equation}\label{nonint_energy}
\mathcal{E}_0(g,\chi)= \frac{1+\chi^2}{2}
- g \frac{\sqrt{|1+\chi|^3}-\sqrt{|1-\chi|^3}}{3} ~.
\end{equation}
Here the first term is the kinetic energy while the second one is the 
spin-orbit coupling contribution.\cite{comment_KE_for_chi_larger_than_one}
For non interacting electrons the value of $\chi$ that minimizes this expression 
depends only on the strength of the spin-orbit coupling and the electron density and, remarkably,  
is in fact uniquely determined by the parameter 
$g$ [defined above in Eq.~(\ref{g_def})] as follows:
\begin{eqnarray}\label{chi_nonint}
\chi_{0}(g)=
\left\{
\begin{array}{cl}
g \sqrt{1-\frac{g^2}{4}}    &\quad {\rm for}~0\leq g <\sqrt{2} ~,\\
\frac{g^2}{4}+\frac{1}{g^2} &\quad {\rm for}~g \geq \sqrt{2} ~.
\end{array}
\right.
\end{eqnarray}

\begin{figure}
\begin{center}
\raisebox{0.7cm}[3cm][0pt]{
\makebox[3.5cm][l]{\includegraphics[width=0.15\textwidth]{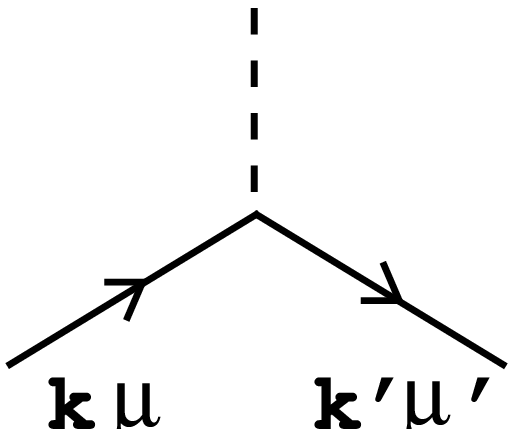}}}
\includegraphics[width=0.2\textwidth]{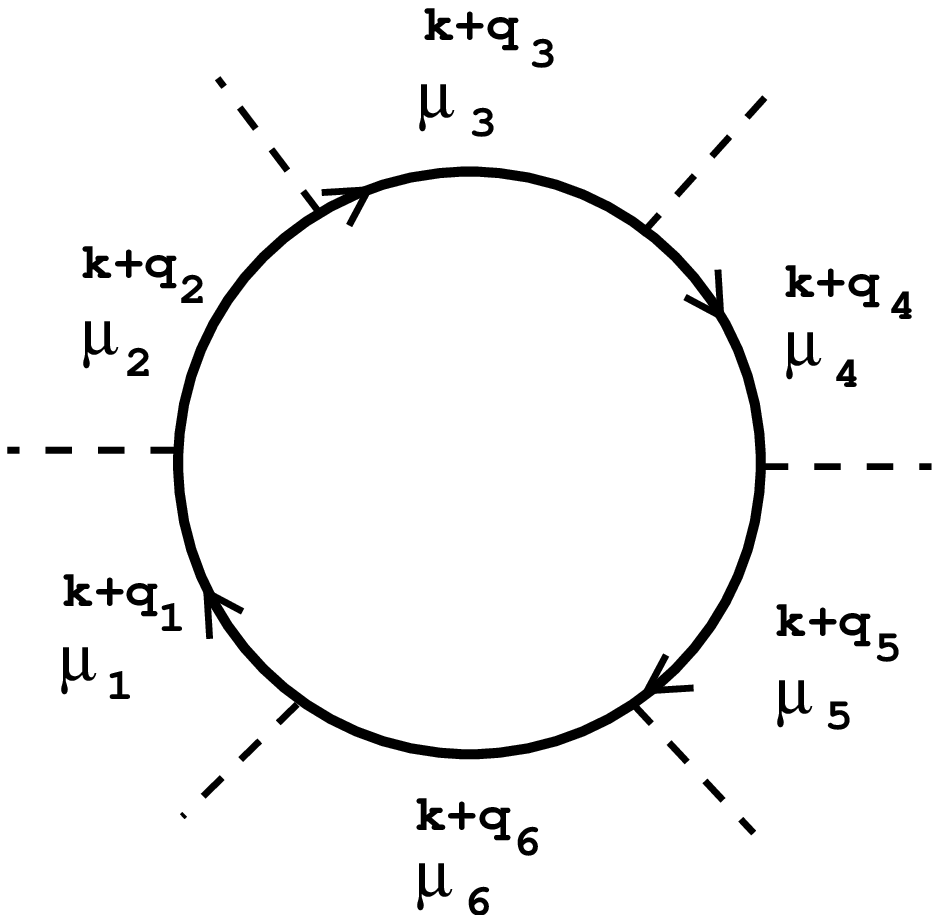}
\caption{\label{diagram_figure} Left: interaction vertex, associated with a  
$\langle {\bf k} \mu | {\bf k}' \mu' \rangle $ factor. Right: 
example of a fermionic loop.}
\end{center}
\end{figure}

The perturbative analysis is very similar to the standard
case without spin-orbit coupling.\cite{TheBook} The fully interacting Green's function 
$ G_\mu(k,t)=-i \langle T \hat b_{{\bf k}\mu}(t)\hat b^\dag_{{\bf k} \mu}(0) \rangle$
($\mu = \pm$)
is obtained as an average of the time ordered product of creation (destruction) operators 
$\hat b^\dag_{{\bf k} \mu}$ ($\hat b_{{\bf k}\mu}$) of the chiral states 
of Eq.~(\ref{phi0kpm}).\cite{comment_Gdiagonal} 
The only modification to the diagrammatic rules is that each vertex, beside the delta 
functions ensuring frequency and momentum conservation, is associated with the scalar product
$\langle {\bf k} \mu | {\bf k}' \mu' \rangle $ [the spinors having been defined in
Eq.~(\ref{phi0kpm})]. This is exemplified in Fig.~\ref{diagram_figure}.
Accordingly, the overall dependence of the diagram  on the variables $\chi$ and $g$ is solely
contained in the non interacting Green's functions. 
The Fourier transform of the latter is given by
\begin{equation}\label{G0}
G_{0\mu}( k,\omega)=\frac{1-n_\mu(k)}{\omega-\epsilon_\mu(k)+i\eta}+
\frac{n_\mu(k)}{\omega-\epsilon_\mu(k)-i\eta} ~,
\end{equation}
where $n_\pm(k)$ is defined in (\ref{n_plus_minus}) and $\eta=0^+$ is an infinitesimal quantity. Notice that the dependence
on $\chi$ is fully contained in the occupation numbers, while the dependence on $g$ 
is only contained in $\epsilon_\mu(k)$. 
A similar set of diagrammatic rules applies to the self-energy, defined through the relation
$G_\mu(k,\omega)=1/\left(\omega-\epsilon_\mu(k)+ \Sigma_\mu(k,\omega)/\hbar\right)$.

\section{An exact property of the energy expansion}\label{energy_expansion}

For any given value of the chirality, the fully interacting total energy per particle 
can be obtained from the following integration over the coupling constant 
formula:\cite{TheBook}
\begin{eqnarray}
\label{total_energy_int_coupl_const}
\mathcal{E}(g,\chi)=\mathcal{E}_0(g,\chi)+\hspace{4cm} \\
+\pi \sum_\mu \int_0^1 \frac{{\rm d}\lambda}{\lambda}
\int \frac{{\rm d}{\bf k}}{(2\pi)^2} \int \frac{{\rm d}\omega}{2\pi}
G^\lambda_\mu(k,\omega)\Sigma^\lambda_\mu(k,\omega)~, \nonumber
\end{eqnarray}
where the non interacting contribution was given in Eq.~(\ref{nonint_energy}). 
In the second term the task is to integrate over $\lambda$ a sum of closed diagrams, 
each multiplied by a factor of $\lambda^{n-1}$, where $n$ is the order of the 
interaction. Consider now the expression $D$ of one of these closed diagrams, contributing to the second term of
Eq.~(\ref{total_energy_int_coupl_const}). It is the purpose of this Section to show that
\begin{equation}\label{derivatives_rel_D}
\left.\frac{\partial^2 D}{\partial^2\chi}\right|_0 = 
\left.\frac{\partial^2 D}{\partial^2 g}\right|_0 =
-\left.\frac{\partial^2 D}{\partial \chi \partial g}\right|_0  ~,
\end{equation}
where the derivatives are evaluated at $\chi=g=0$.
Our treatment is carried out to all orders of perturbation theory, for a generic diagram $D$ and
a general two-body potential $v(q)$. 

To prove Eq.~(\ref{derivatives_rel_D}), we begin by noticing that all the closed diagrams are
in general comprised of several fermionic loops. Let us focus our attention on the contribution to Eq.~(\ref{derivatives_rel_D}) stemming from
of one such loop (schematically illustrated in Fig.~\ref{diagram_figure}). 
If the loop contains $N$ solid lines we can write: 
\begin{eqnarray}
\label{D}
D=\int[\ldots]\sum_{\{\mu_i \}} \int \frac{{\rm d}{\bf k}}{(2\pi)^2}
\int \frac{{\rm d}\omega}{2\pi}
\prod_{i=1}^N  G_{0\mu_i}(p_i,\omega_i) \nonumber
\\
\times ~ ~~\langle {\bf p}_i \mu_i | {\bf p}_{i+1} \mu_{i+1} \rangle  ~,
\end{eqnarray}
where $\{\mu_i \}\equiv \{\mu_1,\mu_2,\ldots \mu_N\}$ and $N+1\equiv 1$. As in 
Fig.~\ref{diagram_figure}, the internal momenta and frequencies are 
${\bf p}_i = {\bf k}+{\bf q}_i$ and $\omega_i = \omega + \Omega_i$ where 
${\bf k}$ and $\omega$ are the momentum and the frequency flowing in the loop. 
Finally the square bracket $[\ldots]$ represents the remaining expression of 
the diagram, which is independent of ${\bf k}$ and $\omega$ and does not enter the
derivation below.

We proceed by examining the first derivatives of Eq.~(\ref{D}) with respect to 
the variables $\chi$ and $g$ for $g=\chi=0$. It is readily seen that $D$ depends 
on $\chi$ and $g$ only through the non interacting Green's functions (\ref{G0}). 
As a consequence, a direct inspection of Eq.~(13) leads to the conclusion 
that both first derivatives for $g = \chi = 0$ are given by a sum of $N$ 
terms, of which the $j$-th term contains the first derivative of the 
corresponding Green's function in the diagram. 
Since both $\left. \frac{\partial
G_{0\mu_j}}{\partial g}\right|_0$ and
$\left. \frac{\partial G_{0\mu_j}}{\partial \chi}\right|_0$
are proportional to $\mu_j$, this yields the factor
\begin{equation}
\sum_{\{ \mu_i \}}  \mu_j \prod_{i=1}^N \langle {\bf p}_i \mu_i | {\bf
p}_{i+1} \mu_{i+1} \rangle = 0,
\end{equation}
which, as indicated, vanishes upon spin summation.

The second derivatives of $D$ have more complicated expressions that are in 
general non-vanishing also for $\chi=g=0$. We observe next that the dependence of
$G_{0\mu}$ on $\chi$ and $g$ is such that $\frac{\partial}{\partial \chi}$ only acts 
on the occupation numbers $n_\pm(k)$, while $\frac{\partial}{\partial g}$ only acts on 
the energy denominators of this function. To prove Eq.~(\ref{derivatives_rel_D}), 
we then introduce as a formal device the quantity
\begin{eqnarray}\label{DD}
D_{a b}  = 2\int[\ldots] \int
\int \frac{{\rm d}\omega}{2\pi }
\frac{{\rm d}{\bf k}_\chi {\rm d}{\bf k}_g}{(2\pi)^2} \,
\, \delta({\bf k}_\chi-{\bf k}_g)
\qquad\qquad  \\
\times \left[\frac{
\vec \nabla_{{\bf k}_a} \cdot \vec \nabla_{{\bf k}_b}}{4} \right]
\prod_{i=1}^N \tilde {\cal G}_{0}(p_{\chi,i},p_{g,i},\omega_i)  ~, \nonumber
\end{eqnarray}
where each of the the indexes $a$, $b$ can be $\chi$ or $g$ while
${\bf p}_{\chi,i}={\bf k}_{\chi}+{\bf q}_i$ and ${\bf p}_{g,i}={\bf k}_{g}+{\bf q}_i$.
This expression corresponds to the same diagrammatic contribution of Eq.~(\ref{D}). The square parenthesis in the first line coincides with the expression omitted in Eq.~(\ref{D}), but is here evaluated at $g=\chi=0$. We have also defined the `resolved' non-interacting Green's function for vanishing spin-orbit coupling ($g=\chi=0$), i.e.
\begin{equation}\label{G0tilde}
\tilde {\cal G}_{0}(k_\chi,k_g,\omega)=\frac{1-n_0(k_\chi)}{\omega-k_g^2+i\eta}+
\frac{n_0(k_\chi)}{\omega-k_g^2-i\eta} ~,
\end{equation}
where $n_0(k)=\theta(1-k)$.
In this expression, the momenta $k_\chi$ and $k_g$ respectively appearing in the occupation 
numbers and the energy denominators are treated as two independent variables.
Notice that, except for the differential operator inside the square parenthesis,
Eq.~(\ref{DD}) coincides with the $\chi=g=0$ expression of the diagram $D$. 
At this point, by integrating by parts (so that the derivatives act on the 
two dimensional delta function), one can see that 
\begin{equation}
\label{derivatives_vs_Dab}
D_{\chi\chi} = D_{gg}=-D_{\chi g} ~.
\end{equation}
The desired identity Eq.~(\ref{derivatives_rel_D}) follows from the relation
\begin{equation}\label{Drelation}
\left.\frac{\partial^2 D}{\partial a \partial b}\right|_0 = D_{ab} ~,
\end{equation}
which is proved next.

To derive Eq.~(\ref{Drelation}) we consider first the equivalence of 
$\left.\frac{\partial^2 D}{\partial \chi^2}\right|_0$ with $D_{\chi\chi}$. 
The former is the sum of several terms, of which some contain the second 
derivative of a single Green's function, and others contain the product
of first derivatives of two distinct Green's functions. The same holds true for $D_{\chi\chi}$
[see Eq.~(\ref{DD})]. The euqlity between corresponding second derivative terms
of $\left.\frac{\partial^2 D} {\partial \chi^2}\right|_0$ and $D_{\chi\chi}$ can 
be established by using the relation
\begin{equation}
\label{derivGG_chi2} 
\left.
\frac{\vec{\nabla}_{{\bf k}_\chi}^2}{4}
~\tilde {\cal G}_0(p_\chi,p_g,\omega)
\right|_{p_\chi = p_g = p } =
\left.
\frac{\partial^2 G_{0\mu}(p,\omega)}{\partial \chi^2}
\right|_0  ~,
\end{equation}
where on the left side we set $p_\chi=p_g=p$ and on the right side $g=\chi=0$. The necessary
factor of $2$ is obtained performing the spin summation
\begin{equation}
\label{spinsum00}
\sum_{\{\mu_i\}} \prod_{i=1}^{N} 
\langle {\bf p}_i \mu_i | {\bf p}_{i+1} \mu_{i+1} \rangle = 2 ~.
\end{equation}
For the first derivative terms, we can use the formula ($a=g, \chi$)
\begin{equation}
\label{derivGG_1} 
\left.
\frac{\vec{\nabla}_{{\bf k}_a}}{2}~\tilde {\cal G}_0(p_\chi,p_g,\omega)
\right|_{p_\chi = p_g =p} =
\, -\left.
\frac{\partial G_{0\mu}(p,\omega)}{\partial a}
\right|_0 \, \frac{\mu {\bf p}}{p}  ~.
\end{equation}
The factors $\frac{{\bf p}_j \cdot {\bf p}_k}{p_j p_k}$ are recovered by performing the
following spin summations
\begin{equation}
\label{spinsum22}
\sum_{\{\mu_i\}} \mu_j \mu_k \prod_{i=1}^{N} 
\langle {\bf p}_i \mu_i | {\bf p}_{i+1} \mu_{i+1} \rangle =
2\frac{{\bf p}_j \cdot {\bf p}_k}{p_j p_k}~.
\end{equation}
This establishes the desired result $\left.\frac{\partial^2 D}{\partial \chi^2}\right|_0=
D_{\chi\chi}$.

The equality $\left.\frac{\partial^2 D}{\partial \chi \partial g}\right|_0=
D_{\chi g}$ can be proved in a similar way. For this case, the following relation proves useful 
\begin{equation}
\label{derivGG_chig} 
\left.
\frac{\vec{\nabla}_{{\bf k}_\chi}\cdot\vec{\nabla}_{{\bf k}_g}}{4}
~\tilde {\cal G}_0(p_\chi,p_g,\omega)
\right|_{p_\chi = p_g =p} =
\left.
\frac{\partial^2 G_{0\mu}(p,\omega)}{\partial \chi \, \partial g}
\right|_0  ~.
\end{equation}

Finally, to prove the equivalence of $\left.\frac{\partial^2 D}{\partial g^2}\right|_0$ 
and $D_{gg}$, we use the following formula
\begin{equation}
\label{derivGG_g2} 
\left.
\frac{\vec{\nabla}_{{\bf k}_g}^2}{4}
~\tilde {\cal G}_0(p_\chi,p_g,\omega)
\right|_{p_\chi = p_g =p} =
\left.
\frac{\partial^2 G_{0\mu}(p,\omega)}{\partial g^2 }\right|_0
+ [{\cal G}_0(p,\omega)]^2 ~,
\end{equation}
where ${\cal G}_{0}(p,\omega)$ is the usual non interacting Green's function 
without spin-orbit coupling ($g=\chi=0$).\cite{TheBook} 
A slight complication arises, due to the $[{\cal G}_0(p,\omega)]^2$ term of (\ref{derivGG_g2}).
This produces an additional contribution of the following form
\begin{eqnarray}\label{deltaDDgg}
&& 2 \int[\ldots] 
\int \frac{{\rm d}\omega}{2\pi }
\int \frac{{\rm d}{\bf k}}{(2\pi)^2} \, 
 \sum_{j=1}^N [{\cal G}_0(p_j,\omega_j)]^2\prod_{i\neq j}  {\cal G}_{0}(p_i,\omega_i)  \nonumber \\
&=&-2 \int[\ldots]
\int \frac{{\rm d}{\bf k}}{(2\pi)^2} 
\int \frac{{\rm d}\omega}{2\pi }  \,
\frac{\partial}{\partial \omega}  \,
 \prod_{i=1}^N  {\cal G}_{0}(p_i,\omega + \Omega_i)  ~. 
\end{eqnarray}
However, (\ref{deltaDDgg}) is seen to vanish upon integration over the loop frequency $\omega$.

\section{An exact property of the self-energy expansion}\label{self_energy}

The fully interacting self-energy $\Sigma_\mu(k,\omega)$ in the presence of spin-orbit coupling
satisfies a similar exact relationship to linear order in $g$. Again, we consider its diagrammatic 
expansion and show that for any given diagram $D_{\Sigma_\mu}$:
\begin{equation}\label{self-energy}
\left. \frac{\partial D_{\Sigma_\mu}(k,\omega)}{\partial g}\right|_0 =
-\frac{\mu}{2}  \frac{\partial D_{\Sigma_0}(k,\omega)}{\partial k}  ~.
\end{equation}
where $\Sigma_0(k,\omega)$ is the interacting self energy without spin-orbit coupling (therefore, $D_{\Sigma_0}$ is obtained setting $g=\chi=0$ in $D_{\Sigma_\mu}$). We will derive Eq.~(\ref{self-energy}) assuming that $\chi \simeq g$ and, as in Eq.~(\ref{derivatives_rel_D}), the left side derivative is evaluated at $g=0$. Within the notation of the previous Section, $D_{\Sigma_\mu}$ is written as
\begin{equation}\label{Dsigma}
D_{\Sigma_\mu} =\int[\ldots]\sum_{\{\mu_i \}} 
\prod_{i=1}^N  G_{0\mu_i}(p_i,\omega_i) 
\prod_{j=0}^N \langle {\bf p}_j \mu_j | {\bf p}_{j+1} \mu_{j+1} \rangle  ~.
\end{equation}
Here, only Green's functions connected which the external momentum ${\bf k}$
are explicitly written. If $1\leq i \leq N$ we set ${\bf p}_i={\bf k}+{\bf q}_i$ while
${\bf p}_0={\bf p}_{N+1}={\bf k}$ and $\mu_0=\mu_{N+1}=\mu$. 

Also in this case the derivative $\frac{\partial}{\partial g}$ involves all the non 
interacting Green's functions appearing in the diagram. On the other hand, as we have discussed
in the previous Section, one needs not to worry about the (vanishing) contribution
of all the sets of Green's functions that are involved in fermionic loops.

Therefore, the expression of $\left. \frac{\partial D_{\Sigma_\mu}}{\partial g}\right|_0$ 
consists of the sum of $N$ terms, involving the first derivatives of the Green's 
functions explicitly appearing in Eq.~(\ref{Dsigma}). The final result is 
\begin{eqnarray}\label{deriv_Dsigma}
\left. \frac{\partial D_{\Sigma_\mu}}{\partial g}\right|_0 =\int[\ldots] \sum_{j=1}^N 
\left( \mu_j \,  \left. \frac{\partial G_{0\mu_j}(p_j,\omega_j)}{\partial g} \right|_0 \right) 
\mu \frac{ {\bf k}\cdot {\bf p}_j}{k \, p_j}\qquad \\
\times {\cal G}_{0}(p_1,\omega_1){\cal G}_{0}(p_2,\omega_2)\ldots 
\widehat{{\cal G}_{0}(p_j,\omega_j)}\ldots {\cal G}_{0}(p_N,\omega_N)~,\nonumber
\end{eqnarray}
where the factor ${\cal G}_{0}(p_j,\omega_j)$ is omitted in the second line. 
We also notice that, in the first line, the expression in the round brackets is independent 
of $\mu_j$ (since $\left. \frac{\partial G_{0\mu}}{\partial g}\right|_0 \propto \mu$).
Furthermore the sums over the spin indexes $\{\mu_i\}\equiv \{\mu_1,\mu_2,\ldots \mu_N\}$ 
were performed using the relation
\begin{equation}\label{spinsumSelfEnergy}
\sum_{\{\mu_i\}} \mu_j \prod_{i=0}^N \langle {\bf p}_i \mu_i | 
{\bf p}_{i+1} \mu_{i+1} \rangle = \mu \frac{{\bf k}\cdot {\bf p}_j}{k \, p_j} ~.
\end{equation}

We then consider the expression $D_{\Sigma_0}$ of the same diagram occurring in the expansion of the
interacting self energy in the absence of spin-orbit coupling:
\begin{equation}\label{Dsigma0}
D_{\Sigma_0} ~=~ \int[\ldots] \prod_{i=1}^N  {\cal G}_{0}(p_i,\omega_i)~.
\end{equation}
Its derivative can be directly compared to Eq.~(\ref{deriv_Dsigma}) by making use of
the relation
\begin{equation}\label{derivG0}
 \frac{\partial {\cal G}_0(p_j,\omega_j)}{\partial k}=
-2\mu_j \,  \left. \frac{\partial G_{0\mu_j}(p_j,\omega_j)}{\partial g}\right|_0 \,\frac{ {\bf k}\cdot {\bf p}_j}{k \, p_j} ~,
\end{equation}
which is easily verified for $\chi\simeq g$. Eq.~(\ref{self-energy}) then immediately follows.

\section{Discussion}\label{discussion}

We have derived by diagrammatic means two exact relationships involving the 
total energy expansion and the self-energy of a two dimensional system in the presence of 
Rashba or Dresselhaus spin-orbit coupling. The first identity is given 
in Eq.~(\ref{derivatives_rel_D}) and involves the second derivatives of the energy diagrams with 
respect to the density dependent dimensionless spin-orbit coupling $g$, 
defined in Eq.~(\ref{g_def}), and the chirality $\chi$, defined in Eq.~(\ref{chi_def}).
Notice that, as made clear in Eq.~(\ref{kpm}), $\chi$ takes into account the renormalized momentum space
repopulation brought about by interactions.\cite{chesi07b} Since Eq.~(\ref{derivatives_rel_D}) is valid for
any diagrammatic contribution, we can formally sum the perturbative expansion in the electron interaction. 
This leads to the following formula, for the small $g$ and $\chi$ expression of the 
extra contribution to the electronic exchange and the correlation energy due 
to the spin-orbit coupling: 
\begin{equation}\label{E_change_g_chi}
\delta \mathcal{E}_{xc}(g,\chi) = C (g-\chi)^2 + \ldots  ~, 
\end{equation}
where the constant $C$ is unspecified, and terms of higher order in $g$ and $\chi$
are omitted. It is clear from this result that the total energy is an extremum
for $\chi \simeq g$. Comparing with Eq.~(\ref{chi_nonint}) we conclude that to 
linear order in $g$ the interactions lead to no change in $\chi$, i.e.
no repopulation, and there is no energy correction $\delta \mathcal{E}_{xc}(g,\chi)$ to quadratic order in $g$.\cite{Abedinpour2010} This remarkable result does not hold for other types of spin-orbit coupling.\cite{chesi07b}

Our second result, Eq.~(\ref{self-energy}), involves the self-energy and allows us to 
write that to linear order in $g$
\begin{equation}\label{self-energy2}
\Sigma_\mu(k,\omega) =
\Sigma_0(k-\frac{\mu g}{2},\omega)+ \ldots ~.
\end{equation}
From this relation we conclude that, to linear order in $g$, all the quasiparticles properties
(effective mass, lifetime, \ldots) on the Fermi surfaces $k_\pm \simeq 1\pm \frac{g}{2}$ 
can be simply obtained from the case without spin-orbit coupling. 
Equation~(\ref{self-energy2}) extends the results of Ref.~\onlinecite{saraga05} where 
this statement was derived in the RPA approximation for Coulomb interactions.
We observe that, similarly to the case of the phonon self-energy in a Fermi liquid, 
the interaction self-energy for the case of spin-orbit coupling is seen to ``ride" the 
Fermi surface. 

Finally, we would like to comment about the relation of our calculation to non-analyticities, which are known  for a long time to occur in the theory of the electron liquid\cite{GellMann1957,Rajagopal1977} and have attracted much attention in more recent years (see, for example, Refs. \onlinecite{Belitz1997,Chubukov2003,Galitski2004,Chubukov2005,Maslov2006,Maslov2009,zak10}). Non-analytic terms in the interactions are fully included in our results, since our derivation applies to all orders in perturbation theory. However, an analytic expansion in the spin-orbit coupling parameters, $g$ and $\chi$, is assumed.

This procedure is justified here because we restrict ourselves to low orders in the spin-orbit coupling parameters (first-order for the self-energy and second order for the total energy). For the total energy, non-analytic corrections can only appear starting from third order in the relevant energy scale, as it can be concluded from a power-counting argument.\cite{Maslov2006,Maslov2009} For example, a contribution to the total energy $\propto |\alpha|E_Z^2$ (where $E_Z$ is the Zeeman energy) is responsible of a non-analytic correction $\propto |\alpha|$ to the spin-susceptibility.\cite{zak10} This contribution to the energy is vanishing in the absence of magnetic field but a term $\propto |\alpha|^3\ln^2 |\alpha| $ can be derived at fourth order in the electron interaction.\cite{zak_unpubl} A non-analytic correction to the exchange energy is also obtained in Ref.~\onlinecite{Chesi2010}, proportional to $\chi^4 \ln|\chi|$. For the self-energy, no terms linear in $\alpha$ were found in Refs.~\onlinecite{saraga05,Agarwal2010} and non-analytic corrections proportional to $\alpha^2 \ln|\alpha|$ were derived in Ref.~\onlinecite{saraga05}. All these non-analyticities do not affect our main results, Eqs.~(\ref{E_change_g_chi}) and (\ref{self-energy2}), since they only appear at higher order in the spin-orbit coupling. 

\begin{acknowledgments}
We would like to thank Dmitrii L. Maslov and Robert A. \.Zak for helpful discussions and Yuli Lyanda-Geller for pointing us to Ref.~\onlinecite{aleiner01}.  SC acknowledges support by
NCCR Nanoscience, Swiss NSF, and CIFAR.
 
\end{acknowledgments}


\begin{thebibliography}{16}
\bibitem[{com({\natexlab{a}})}]{comment_genchi}
\bibinfo{note}{
On the other hand, a value $\chi \geq 1$ corresponds to a low density
  case in which only the lowest chirality band is occupied, the occupied states
  forming a ring in momentum space. In this case the regular chirality is simply 1, and differs from the value of $\chi$.
}
\bibitem[{com({\natexlab{b}})}]{comment_KE_for_chi_larger_than_one}
\bibinfo{note}{The expression for the case of $\chi \geq 1$ is simply obtained
  by substituting $2\chi$ for $1+\chi^2$ in the numerator of the kinetic energy
  term.}
\bibitem[{com({\natexlab{c}})}]{comment_Gdiagonal}
\bibinfo{note}{For the paramagnetic state, the Green's function is diagonal in
  $\mu$. We assume $\bf k$ in the $x$ direction and consider the reflection
  operator $\hat{R}_y$, corresponding to $y \to -y$, which gives $\langle
  \hat{R}_y^{-1} \hat b_{{\bf k}\nu} b^\dag_{{\bf k}\mu} \hat{R}_y\rangle =\mu
  \nu \, \langle \hat b_{{\bf k}\nu} \hat b^\dag_{{\bf k}\mu}\rangle$. If the
  ground state has a definite parity with respect to $\hat{R}_y$, the
  expectation value should be invariant, and the $\mu\nu=-1$ case gives zero.}
\end{thebibliography}

\end{document}